# Current-driven Magnetization Reversal and Spin Wave Excitations in Co/Cu/Co Pillars

J. A. Katine, F. J. Albert, R. A. Buhrman
*School of Applied and Engineering Physics, Cornell University, Ithaca, NY 14853*

E. B. Myers, D. C. Ralph
*Laboratory of Atomic and Solid State Physics, Cornell University, Ithaca, NY 14853*

Using thin film pillars ∼100 nm in diameter, containing two ferromagnetic Co layers of different thicknesses separated by a paramagnetic Cu spacer, we examine effects of torques due to spin-polarized currents flowing perpendicular to the layers. In accordance with spin-transfer theory, spin-polarized electrons flowing from the thin to the thick Co layer can switch the magnetic moments of the layers antiparallel, while a reversed electron flow causes switching to a parallel state. When large magnetic fields are applied, the current no longer fully reverses the magnetic moment, but instead stimulates spin-wave excitations.

Theoretical calculations have recently demonstrated that when a spin-polarized current passes through a ferromagnetic conductor, the transfer of angular momentum from the polarized current exerts a torque on the magnetic moment of the conductor. At sufficiently high current densities, $J$, this interaction has been predicted to stimulate spin wave excitations [1–3] or to possibly flip the magnetic moment of an individual domain [2]. Evidence for spin-transfer-induced excitations has been reported previously in point-contact measurements of Cu/Co multilayers [4,5] and nickel wires [6], and spin-transfer-driven moment reversals have been reported in manganite junctions [7] and point contacts [5]. The point contact studies [4,5] involved continuous ferromagnetic layers in which extreme ($\sim 10^9$ A/cm$^2$) current densities were required to reverse the magnetization direction of a localized domain. In this Letter, we study the effects of spin-transfer in a simpler geometry – lithographically patterned pillars consisting of two Co layers of different thicknesses separated by a paramagnetic Cu layer spacer. Since the pillar devices are a well-controlled geometry with a known thickness and diameter, they facilitate a quantitative study of the spin-transfer phenomena and tests of the theoretical models. With these devices we confirm the results of [5] that, in low applied magnetic fields $H$, spin-polarized electrons flowing from the thin Co layer to the thick layer can switch the moment of the thin layer antiparallel to the thick-layer moment, while a reversed current produces a switch back to the parallel orientation. However, in the pillar geometry, the spin-polarized current is incident on an isolated ferromagnetic particle; hence, the domain which reverses is not exchange coupled to a continuous magnetic layer. The reorientations therefore occur at far more modest current densities ($<10^8$ A/cm$^2$) than in [4,5]. In large $H$, the spin-transfer effect cannot produce a full reversal of the thin-layer moment, but gives instead a precessing spin-wave excitation.

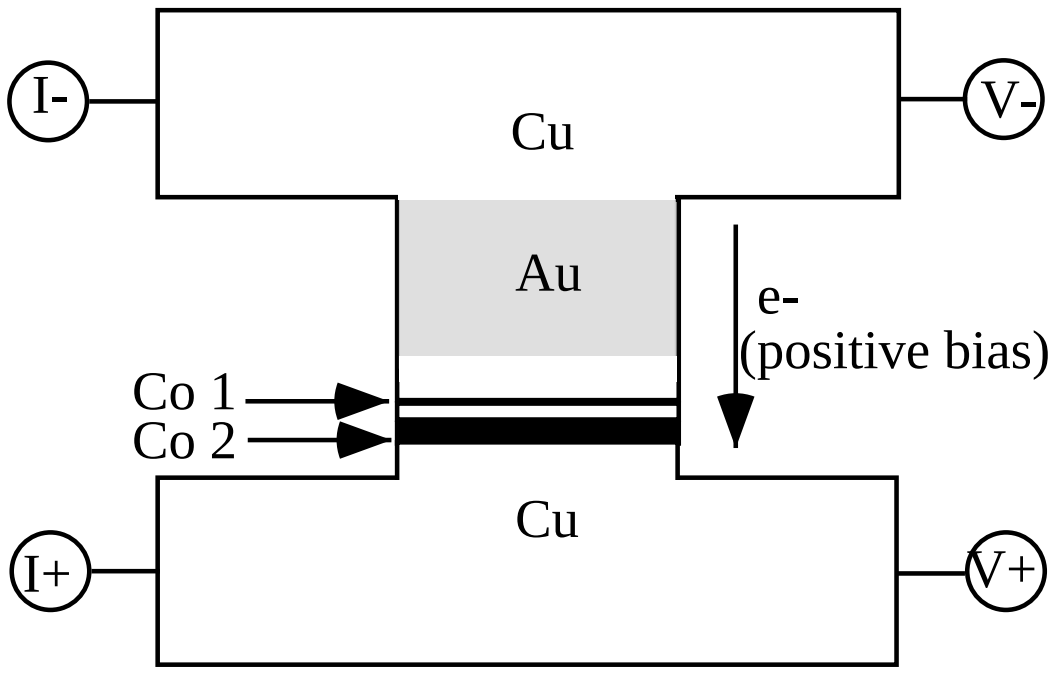

FIG. 1. Schematic of pillar device with Co (dark) layers separated by a 60 Å Cu (light) layer. At positive bias, electrons flow from the thin (1) to the thick (2) Co layer.

Figure 1 is a schematic representation of the pillar device. A number of groups [8–11] have used lithographic patterning to perform CPP (current perpendicular-to-plane) measurements on magnetic multilayers exhibiting giant magnetoresistance (GMR). Here we use GMR as a probe of the relative orientation of the two Co layers, in pillars much narrower than those previously made. We begin fabrication by sputtering 1200 Å Cu/100 Å Co/60 Å Cu/ 25 Å Co/ 150 Å Cu/ 30 Å Pt/ 600 Å Au onto an oxidized Si substrate. The difference in thickness for the Co layers allows the magnetization direction of the thicker layer to be held fixed, so that the polarity of the current bias associated with the spin-transfer excitations in the thinner layer can be determined. Electron beam lithography, evaporation and lift-off are used to pattern ∼100 nm diameter, 500 Å thick Cr dots on top of the sputtered film. These serve as a mask during an ion milling step that etches through the bottom Co layer. The final device diameter is 130 $\pm$ 30 nm. Following planarization with polyimide, reactive ion etching and photolithography are used to uncover the Au surface of the pillar, and pattern the top leads.

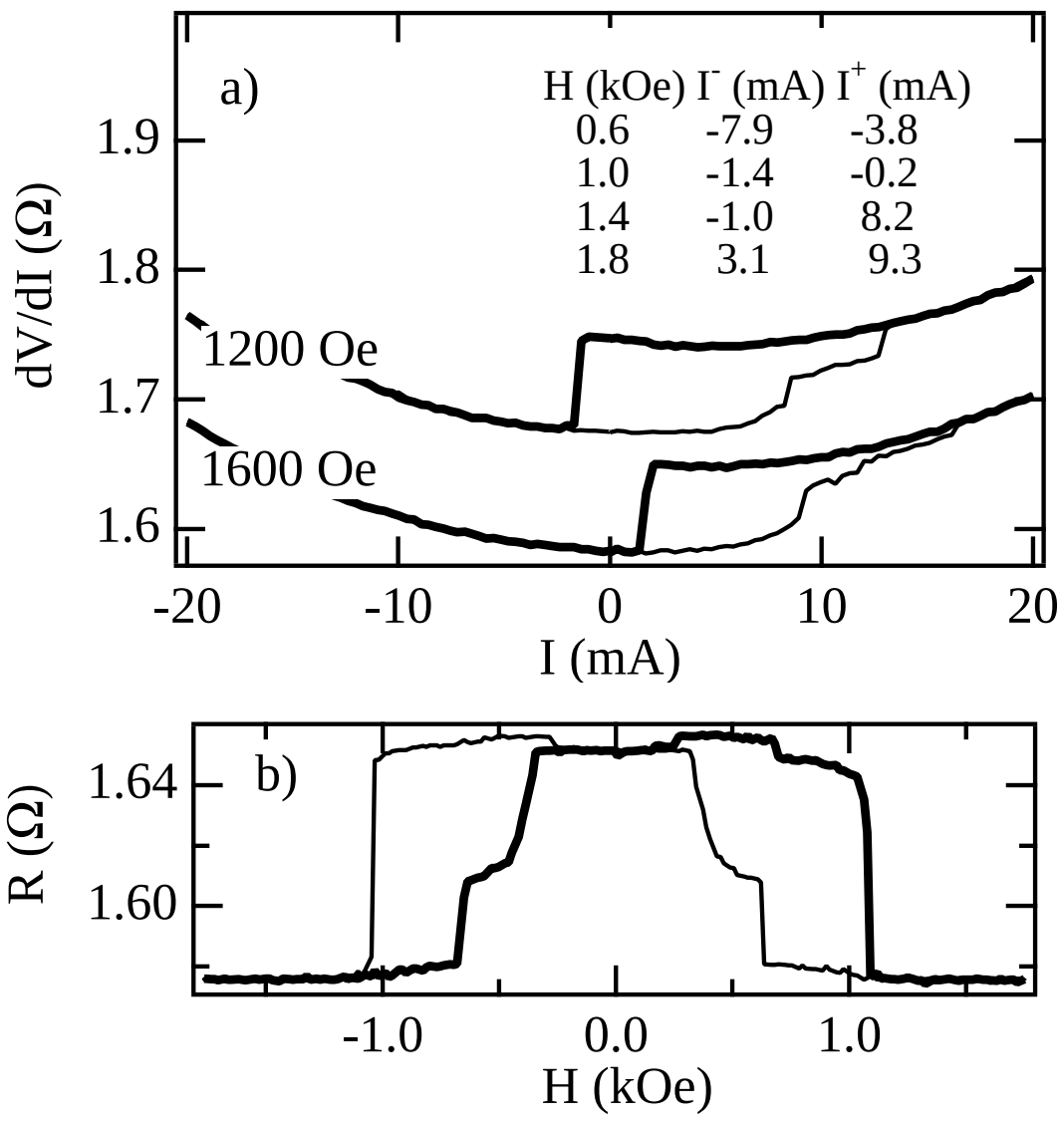

FIG. 2. (a) $dV/dI$ of a pillar device exhibits hysteretic jumps as the current is swept. The current sweeps begin at zero; light and dark lines indicate increasing and decreasing current respectively. The traces lie on top of one another at high bias, so the 1200 Oe trace has been offset vertically. The inset table lists the critical currents at which the device begins to depart from the fully parallel configuration ($I^+$) and begins to return to the fully aligned state ($I^-$). (b) Zero-bias magnetoresistive hysteresis loop for the same sample.

Figure 2a shows graphs of the room temperature differential resistance $dV/dI$ vs. $I$ in one pillar, taken with magnetic fields $H$ of 1200 and 1600 Oe applied in the plane of the film. $H$ fixes the magnetization of the thick Co layer, $\vec{M}_2$, and also helps prevent the formation of domains within the layers. $dV/dI$ is measured using lock-in techniques with a 10 $\mu$A a.c. excitation, and the d.c. resistance $R_{\rm dc}$ is monitored simultaneously. Consistent with [5], we define positive bias such that electrons flow from the thin Co layer to the thick Co layer. Examining the 1200 Oe data, we begin for $I$=0 with the device in the low resistance state. $I$ is first swept positive and $dV/dI$ (and $R_{\rm dc}$, not pictured) increases in two discrete jumps at 9 and 13 mA, corresponding to $J$=0.7 and 1.0$\times 10^8$ A/cm$^2$ respectively. The curve is hysteretic, with the device remaining in this high $R$ state until the current is swept to negative values where it returns to the low $R$ orientation in a single jump. In addition to the jumps, there is a gradual rise in $R$ with increasing bias due to growth of electron-magnon and electron-phonon scattering. We attribute the jumps in $R$ to changes in the relative alignment of the magnetization of the Co layers, with the low resistance state reflecting parallel alignment of the layer magnetizations, and the high resistance state corresponding to antiparallel alignment. The 73 m$\Omega$ total difference in $R$ is very close to that observed for magnetization realignment as a function of $H$ at $I$=0, shown in Fig. 2b.

From Fig. 2b, we note that for $I$=0 the sample returns to the high-resistance anti-parallel state before the sign of $H$ is reversed, indicating the presence of antiferromagnetic coupling between the layers. Although exchange coupling through the 60 Å Cu layer should be quite weak, in a narrow pillar geometry antiferromagnetic alignment may be created by the presence of magnetostatic edge charges, evidence of which has been reported in significantly wider multilayer pillars [12]. We calculate that a magnetostatic interaction $H_{\rm ex}$ of roughly 1000 Oe should be felt by the edge of thinner Co layer in our geometry. Since $R$ is only a measure of the relative alignment of the magnetization of the two layers, when a plateau is observed in the magnetoresistance at some intermediate value (*e.g.* at $\pm$ 500 Oe in Fig. 2b), we do not know if it is because a layer is essentially single domain and has had its magnetization rotate into a quasi-stable configuration, or if it is because layer 1 contains two domains with different coercive fields. The nature of the two discrete upward jumps in resistance as a function of $I$ in Fig. 2a is similarly ambiguous.

That a sufficiently large spin-polarized current can flip the magnetization of an isolated particle was predicted by Slonczewski by incorporating spin-transfer effects into the Landau-Lifshitz equation [2]. Adapting Slonczewski's argument to our geometry, consider a simplified model of the pillar devices in which the thin Co layers lie in the a-c plane with the b-axis perpendicular to the thin film. $H$ is directed along the c-axis which is also the axis of a uniaxial anisotropy term of strength $H_{\rm an}$. If we assume that the magnetization of layer 2 is fixed and that layer 1 is a single domain particle of volume $V$ and total spin $\vec{S}$ ($S \equiv |\vec{S}| = MV/\gamma\hbar$, where $M$=1420 emu/cm$^3$ for Co and $\gamma$ is the gyromagnetic ratio), we then have

$$\frac{d\hat{s}}{dt} = \hat{s} \times \left\{ \gamma [H_{\rm eff}\hat{c} - 4\pi M(\hat{s}\cdot\hat{b})\hat{b}] - \alpha\frac{d\hat{s}}{dt} - \frac{Ig(\theta)}{eS}\hat{c}\times\hat{s} \right\}, \quad (1)$$

where $H_{\rm eff}$=$H_{\rm an}\cos(\theta)$ + $H$ − $H_{\rm ex}$ is the sum of the anisotropy, applied, and exchange (but not demagnetizing) fields and $\theta$ represents the angle between the magnetization vectors of layers 1 and 2. The coefficient $\alpha$ is the phenomenological Gilbert damping parameter. The final term incorporates spin-transfer effects, in which $g(\theta)$ is a coefficient that depends on the polarization of the electrons, and is calculated to increase monotonically with $\theta$ [2]. In the absence of the spin-transfer term and damping, the solution of this equation is elliptical precession of $\hat{s}$ about the c-axis with $S_{\rm a} = Acos(\omega t)$, $S_{\rm b} = -A(\gamma H_{\rm eff}/\omega)sin(\omega t)$, and $\omega^2 = \gamma^2 H_{\rm eff}(H_{\rm eff} + 4\pi M)$. Damping causes the amplitude $A$ of the precession to decay with time, while depending on the sign of $I$, the spin-transfer term can amplify or attenuate the precession amplitude. For small amplitudes of precession $A$ about $\theta$=0, the time-averaged rate of change in the total energy is

$$\left\langle \frac{dE}{dt} \right\rangle \approx A^2 \frac{H_{\rm eff} M}{S^2} \Big\{ -\alpha\gamma[H_{\rm eff} + 2\pi M] + Ig(\theta)/(eS) \Big\}. \tag{2}$$

Near $\theta$=0, negative values of $\langle dE/dt\rangle$ indicate a decay of the spin precession toward $\theta$=0. A similar expression may be derived near $\theta=\pi$. If we begin near parallel alignment ($\theta\approx0$) and ramp the current, then that alignment remains stable until $I > I_c^+ = \alpha\gamma eS[H_{\rm eff}(0)+2\pi M]/g(0)$, after which $\theta=\pi$ is the only stable configuration. Conversely, if the applied field is not too large ($|H - H_{\rm ex}| < H_{\rm an} + 2\pi M$) and if the device is in the antiparallel alignment, it will remain there until the current decreases below $I_c^- = \alpha\gamma eS[H_{\rm eff}(\pi) - 2\pi M]/g(\pi)$, when it will switch into the parallel configuration.

This model correctly predicts the symmetries of hysteretic switching observed in our devices and in previous point contact studies [5]. The dependence of the switching on the direction of $I$ is strong evidence that a spin-transfer mechanism and not the Oersted fields created by the current flow is responsible for the effect. Since an applied field favors parallel alignment, the model predicts that increasing $H$ should make both $I_c^-$ and $I_c^+$ more positive. This is indeed the case in our devices as illustrated by the inset table in Fig. 2a where we list, for different values of the applied field, the critical currents $I^+$ at which the device starts to leave the fully parallel state and the currents $I^-$ at which the device begins to return to parallel alignment. Lower values of $H$ are not included since the well-ordered switching behavior shown in Fig. 2a is not present for $H\leq$500 Oe. This is not surprising since the assumption that the thick Co layer is a single domain of fixed magnetization may no longer be valid.

Although qualitatively in accord with the Landau-Lifshitz model, the critical current data are not in quantitative agreement. The values of $H_{\rm an}$ and $H_{\rm ex}$ are not precisely known in our devices, but a comparison can be made to the predicted critical currents by examining how $I^+$ and $I^-$ change relative to $H$. From the inset in Fig. 2a, we see that a change in $H$ of 1 kOe increases $I^+$ by approximately 12 mA, and $I^-$ by 5 mA. Assuming a Co polarization of 38% [13] and a damping coefficient $\alpha$=0.007 determined from ferromagnetic resonance studies of Co [14], the predicted change in $I_c^+$ with $H$ is roughly 0.5 mA/kOe, more than an order of magnitude below the values seen in our devices. Of course, the magnetic transitions of the thin Co layer are, in moderate magnetic fields, likely to be more complicated than the uniform rotation of a single domain particle assumed in the model. In particular the gradual onset of the transition and the multi-plateau features often observed suggest that multiple domains and possibly domain wall motion may be involved. Recent studies have suggested that in the latter case a much larger value of $\alpha$ may apply [15].

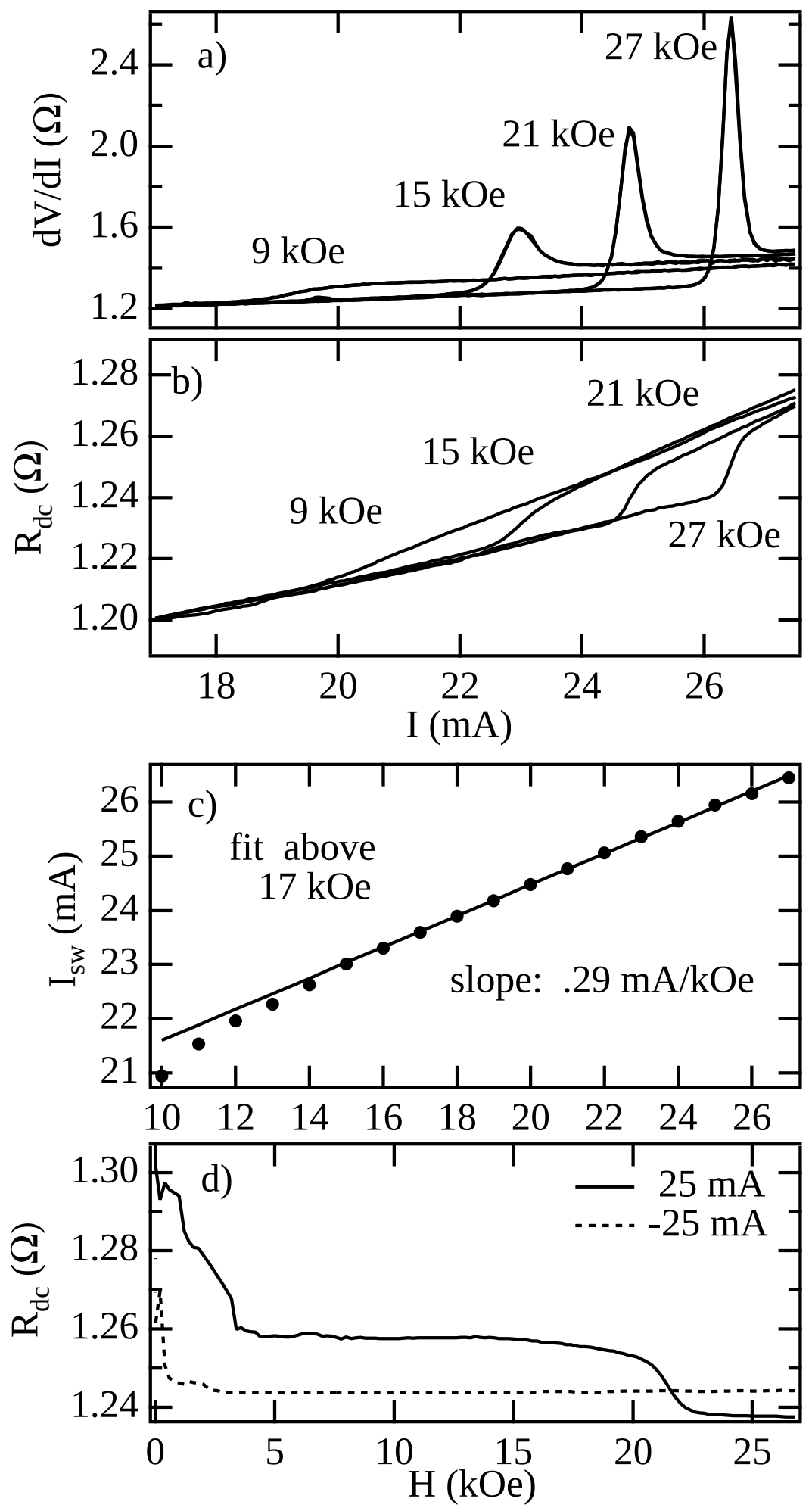

FIG. 3. (a) As $H$ increases, the spike in the $dV/dI$ traces occurs at a larger critical current $I_{\rm sw}$. (b) $R_{\rm dc}$ traces of the same measurements as in (a). (c) Plot of $I_{\rm sw}$ vs. $H$. (d) $R_{\rm dc}$ vs. $H$ with a 25 mA bias current applied in both directions.

As $H$ is increased above 2.3 kOe, discrete switching behavior is no longer observed. Instead spikes in $dV/dI$ emerge as shown in Fig. 3a. Figure 3b shows the corresponding $R_{\rm dc}$ traces. The data presented were taken at 4.2 K, though the spikes were also present at 300 K. Here $H$ was applied in the plane of the film, but similar results were obtained for high $H$ perpendicular to the plane. These data strongly resemble those observed in point contact studies of unbounded multilayer films [4,5]. Figure 3d plots $R_{\rm dc}$ vs. $H$ with constant bias currents of $\pm$25 mA. For the +25 mA curve, the resistance maximum near $H$=0 demonstrates antiparallel alignment, and the low-resistance region $H>$22 kOe is for parallel alignment.

The plateau between 5 and 20 kOe for the +25 mA curve, where the d.c. resistance has an intermediate value, corresponds to the region of increased resistance beyond the peaks in the $dV/dI$ vs. $I$ plots in Fig. 3a. We can therefore argue that the spikes in Fig. 3a do not denote a full reversal of the thin-layer moment, but must correspond to a precessing spin-wave state in between parallel and antiparallel alignment.

To excite the spin wave mode the injected current must provide a torque deflecting the moment of the thin layer away from the aligned configuration. As discussed earlier, this corresponds to a positive bias in our devices. We have studied several dozen samples and always observe the spikes in $dV/dI$ on the positive side. In many instances, there are several spikes present which may be the result of inhomogeneities within the samples. Unlike point contacts where similar spikes were observed previously, the current density in the pillar devices is constant through both layers, which makes clear the intrinsic asymmetry of the phenomenon. On the negative bias side, we have performed measurements out to -80 mA ($J$=6×10$^8$ A/cm$^2$), without ever observing spikes in $dV/dI$.

Two different mechanisms explaining how bias currents may generate spin waves have been proposed. Berger argues that spin waves can occur once $e\Delta\mu = h\nu$, where $\Delta\mu$ is the spin splitting of the chemical potential and $\nu$ is the ferromagnetic resonance frequency [1]. This model is probably not applicable in our devices, since $k_BT \gg h\nu$ at room temperature, so that no sharp spectroscopic signal should exist. Slonczewski [16] has approached the spin-wave excitations using the same spin-transfer framework applied to calculate $I_c^+$. In this approach the existence of a stable spin-precessing mode is somewhat curious, because the solution of the Landau-Lifshitz equation for $H \gg 2\pi M$ does not have stable states other than $\theta$=0 or $\pi$, as long as $g(\theta)$ increases monotonically with $\theta$ and $\alpha$ is a constant independent of $\theta$. However, the assumption that $\alpha$ is constant is the result of a small angle expansion [17]. Keeping next-to-leading-order terms, the argument in [17] predicts $\alpha \propto (1-\cos(\theta))/\sin^2(\theta)$, which is sufficient to stabilize a precessing spin-wave mode at large $H$. Other non-linear effects might also contribute to $\alpha(\theta)$. Although a potential field-independent offset may exist in the pillar geometry, the critical current $I_{\rm sw}$ required to excite the spin-wave mode at large $H$ in the Slonczewski model [16] is otherwise identical to $I_c^+$ calculated above. If we use the value of 0.14 for $g(0)$ calculated using a WKB approximation [2], the slope of 0.29 mA/kOe in Fig. 3c corresponds to $\alpha$=0.005. In sharp contrast to the case for the magnetic reversals, this value agrees quite well with the previously quoted $\alpha$ of 0.007 measured in ferromagnetic resonance experiments on Co films. Conversely, in our previous point contact measurements, where a spin wave was induced in a magnetic region exchange coupled to a unbounded film, damping coefficients 10-50 times larger were necessary to explain how $I_{\rm sw}$ scaled with increasing field [5].

In summary, we have fabricated narrow pillars containing Co/Cu/Co layers. Using the GMR effect as a probe, we have demonstrated that an applied current can be used to controllably flip the relative magnetization alignment of the Co layers, in general accord with spin-transfer theory. For larger $H$, a current bias of the proper polarity can excite uniformly precessing spin wave modes, and in this regime we find excellent quantitative agreement between our experimental data and the spin-transfer theory, particularly with respect to the damping parameter $\alpha$. On the other hand, the effect of the external field on the critical currents required to induce the switching in low $H$ indicates a value of $\alpha$ more than ten times higher. We tentatively attribute this to the switching being more complicated than a simple uniform rotation, as indicated by the signals shown in Fig. 2. This suggests that $\alpha$ varies at the microscopic level depending on the nature of the magnetic phenomena and that spin-transfer experiments can examine such variations.

We thank J. Slonczewski and W. J. Gallagher for helpful discussions. This work was supported in part by DARPA through the Office of Naval Research and by the National Science Foundation through the Cornell Center for Materials Research and through use of the National Nanofabrication Users Network.

[1] L. Berger, Phys. Rev. B **54**, 9353 (1996).
[2] J. Slonczewski, J. Magn. Magn. Mater. **159**, L1 (1996).
[3] Ya. B. Bazaliy, B. A. Jones, and S.-C. Zhang, Phys. Rev. B **57**, R3213 (1998).
[4] M. Tsoi, *et al.*, Phys. Rev. Lett. **80**, 4281 (1998); **81**, 493 (1998) (E).
[5] E. B. Myers, *et al.*, Science **285**, 867 (1999).
[6] J.-E. Wegrowe, *et al.*, Europhys. Lett. **45**, 626 (1999).
[7] J. Z. Sun, J. Magn. Magn. Mater. **202**, 157 (1999).
[8] M. A. M. Gijs, *et al.*, Appl. Phys. Lett. **63**, 111 (1993).
[9] W. Vavra, *et al.*, Appl. Phys. Lett. **66**, 2579 (1995).
[10] K. Bussmann, *et al.*, IEEE Trans. Magn. **34**, 924 (1998).
[11] J.P. Spallas, *et al.*, IEEE Trans. Magn. **33**, 3391 (1997).
[12] T. L. Hylton, *et al.*, Appl. Phys. Lett. **67**, 1154 (1995).
[13] S. K. Upadhyay, *et al.*, Phys. Rev. Lett. **81**, 3247 (1998).
[14] F. Schreiber, *et al.*, Solid State Comm. **93**, 965 (1995).
[15] T. Ono, *et al.*, Science **284**, 468 (1999).
[16] J. Slonczewski, J. Magn. Magn. Mater. **195**, L261 (1999).
[17] H. B. Callen, J. Phys. Chem. Solids **4**, 256 (1958).